\documentstyle[aps,epsf,prb]{revtex}
\tighten
\begin{document}
\draft
\title{Fractionated crystallisation in a polydisperse mixture of hard spheres}
\author{Paul Bartlett}
\address{Department of Chemistry, University of Bath, Bath, BA2 7AY, United 
Kingdom}
\date{September 8, 1998}
\maketitle

\begin{abstract}
We consider the nature of the fluid-solid phase transition in a polydisperse 
mixture of hard spheres. For a sufficiently polydisperse mixture ($\sigma > 
0.085$) crystallisation occurs with simultaneous fractionation. At the 
fluid-solid boundary, a broad fluid diameter distribution is split into a number 
of narrower fractions, each of which then crystallise. The number of crystalline 
phases increases with the overall level of polydispersity. At high densities, 
freezing is followed by a sequence of demixing transitions in the polydisperse 
crystal.

\end{abstract}


\section{Introduction}

Equal-sized hard spheres constitute probably the simplest example of a purely 
entropic material. In a hard-sphere system there is no contribution to the 
internal energy $U$ from interparticle forces so that $U$ is a constant, at a 
fixed temperature. Minimising the free energy, $F=U-TS$, is thus simply 
equivalent to maximising the entropy $S$. Consequently, the structure and phase 
behaviour of hard spheres is determined solely by entropy. Although the 
hard-sphere model was originally introduced as a mathematically simple model of 
atomic liquids\cite{Hansen} recent work has demonstrated it usefulness as a 
basic model for complex fluids\cite{russel}. Suspensions of submicron 
poly(methyl methacrylate) or silica colloids, coated with a thin polymeric layer 
so that strong repulsions dominate the attractive dispersion forces between the 
colloidal cores, behave in many ways as hard spheres. In the last decade, a 
great deal of effort has been devoted to systematic studies of such colloidal 
`hard-sphere' systems\cite{Pusey-853}. For sufficiently monodisperse colloids, 
crystallisation is observed at densities similar to those predicted by computer 
simulation for hard spheres\cite{Pusey-81,Paulin-47}. Measurements of the 
osmotic pressure and compressibility show a similar dramatic agreement with 
predicted hard sphere properties\cite{Pusey-853}. 

There is however one important and unavoidable difference between colloids and 
the classical hard sphere model which is frequently overlooked. Whereas the 
spheres in the classical model are identically-sized ({\em i.e.\/} monodisperse) 
colloidal particles have an inevitable spread of sizes which is most 
conveniently characterised by the polydispersity, $\sigma$, defined as
\begin{equation}
\sigma = \left(\overline{R^{2}} - \bar{R}^{2} \right)^{\frac{1}{2}} / \bar{R}
\end{equation}
where 
\begin{equation}
\rho\overline{R^{n}} = \int dR \rho(R) R^{n} 
\end{equation}
with $\rho(R)$ the density distribution and $\rho = \int dR \rho(R)$.

Recent work has revealed that as soon as a hard sphere suspension is allowed to 
enjoy a significant degree of polydispersity, several interesting new phenomena 
arise. Experiments find that the crystallisation transition is suppressed 
altogether at $\sigma \approx 0.12$ while samples with $\sigma = 0.075$ 
crystallise only slowly in the coexistence region and not above the melting 
transition\cite{Pusey-853}. Other examples of the effects of polydispersity are 
the appearance of a demixing transition in a polydisperse fluid of hard 
spheres\cite{Warren-1239,Cuesta-1264} and the observation of a liquid-vapour 
transition in polydisperse adhesive hard spheres\cite{Sear-1248}.

The effect of polydispersity on the crystallisation of hard sphere colloids has 
been investigated by computer 
simulation\cite{Dickinson-848,Bolhuis-824,Phan-1265}, density 
functional\cite{Barrat-810,McRae-414}, and analytical 
theories\cite{Bartlett-983,Bartlett-1266}. The picture that emerges is 
remarkably consistent. All calculations find that the fluid-solid phase 
transition vanishes for polydispersities above a certain critical level 
$\sigma_{c}$. The phase diagram for small polydispersities ($\sigma \leq 
\sigma_{c}$) has been rationalised \cite{Bartlett-1266} in terms of the 
appearance of an additional high density crystal-to-fluid transition, in a 
polydisperse system. While, at low polydispersities hard spheres display, with 
increasing density, the conventional fluid-to-crystal transition, at higher 
polydispersities re-entrant behaviour is predicted. The two freezing transitions 
converge to a single point in the $(\phi,\sigma)$ plane which is a polydisperse 
analogue of the point of equal concentration found in molecular 
mixtures\cite{Landau-1238}. At this singularity, the free energies of the 
polydisperse fluid and crystal phases are equal. 

The purpose of this note is to examine the fate of a {\em highly\/} polydisperse 
($\sigma > \sigma_{c}$) hard sphere fluid. Previous theoretical research has not 
been able to identify a fluid-solid transition for $\sigma > \sigma_{c}$ so it 
is generally believed that the equilibrium phase is disordered at all densities 
up to close packing. Several years ago, Pusey\cite{Pusey-829} suggested that a 
highly polydisperse suspension might crystallise by splitting the broad overall 
distribution into a number of narrower distributions of polydispersity 
$\sigma_{s}$, each of which could be accommodated within a single crystalline 
phase. Each crystal would then have a correspondingly different mean size with 
the number of crystalline phases increasing with the overall polydispersity. For 
fractionated crystallisation to occur the total free energy of the set of 
multiple crystals must be lower than that of the equivalent polydisperse fluid. 
This can only happen if the reduction in free energy as particles are removed 
from a fluid and placed in a crystal is sufficiently large to exceed the loss of 
entropy of mixing as the distribution is partitioned. This is a delicate balance 
and it is far from obvious where the result lies. Pusey, for instance, 
originally suggested that fractionation would generate crystals with a 
polydispersity of $\sigma_{c}$ so as to minimise the number of crystal phases 
required and the subsequent loss of entropy of mixing. However, as noted above, 
at $\sigma_{c}$ the free energy of the polydisperse fluid and crystal phases are 
equal\cite{Bartlett-1266} and so there is no driving force for fractionation.

In earlier work\cite{Bartlett-1266} the possibility of fractionated 
crystallisation for $\sigma > \sigma_{c}$ was considered but no conditions where 
two crystal phases could coexist could be found. Here we re-examine the 
stability of a polydisperse hard-sphere fluid using a much simpler approach. 
Rather than solving the equations of phase equilibria in a polydisperse system 
we restrict ourselves to the easier task of comparing the free energies, at the 
same density and temperature, of crystal and fluid phases. We find, in agreement 
with Pusey\cite{Pusey-829}, that fractionation occurs in polydisperse hard 
sphere mixtures but that the polydispersity of the resulting crystals is 
substantially less than $\sigma_{c}$.

The rest of the paper is organised as follows: in section \ref{model} we present 
our model for the free energies of the polydisperse fluid and crystal phases. 
The stability diagram is present in section \ref{results}. Finally, in section 
\ref{discussion} we summarise and conclude.

\section{The model}
\label{model}

Our model consists of $N$ hard sphere particles in a volume $V$, at an overall 
density of $\rho = N/V$. Each particle has a diameter $R$ drawn from some 
overall distribution $\rho(R)$ so that $\rho = \int dR \rho(R)$. Previous work 
has suggested\cite{Pusey-829} that the thermodynamic properties of a 
polydisperse system are relatively insensitive to the detailed {\em shape\/} 
assumed for the diameter distribution, at least when the polydispersity is small 
($\sigma < 0.1$). Indeed the phase behaviour has been found\cite{Bartlett}, to a 
rather good approximation, to be a function only of the first three moments of 
the diameter distribution, $\rho$, $\overline{R}$, and the normalised width 
$\sigma$. For a broad diameter distribution, higher moments will presumably also 
need to be considered. Consequently, the phase behaviour will become 
increasingly sensitive to the shape of the distribution assumed. However we 
expect that the width of the distribution will still play an important if not 
the dominant role in determining the generic features of the phase behaviour of 
a polydisperse system. Since our concern here is with establishing these general 
features of the polydisperse phase diagram we have taken $\rho(R)$ as a simple 
rectangular distribution,
\begin{equation}
\rho(R) = \rho \left\{ \begin{array}{ll}
                        0 & R < \bar{R}(1-w/2) \\
                        (\bar{R}w)^{-1} & \bar{R}(1-w/2) \leq R \leq    
\bar{R}(1+w/2) \\
                        0 & R > \bar{R}(1+w/2)
				\end{array}
		  \right.
\label{rectangular}
\end{equation}
where $\bar{R}$ is the mean diameter and the normalised width $w$ is related 
directly to the conventional polydispersity $\sigma$ by $w = 2\sqrt{3} \sigma$. 
It will prove convenient to choose as variables the mean diameter $\bar{R}$, the 
polydispersity $\sigma$ and the volume fraction $\phi = 
(\pi/6)\int{dR\rho(R)R^{3}} = (\pi/6)\rho \bar{R}^{3}(1+w^{2}/4)$. 

Our first task is to calculate the thermodynamics of the fluid state. The free 
energy density ($f = F/V$) may be conveniently split into ideal and excess 
portions
\begin{equation}
f = f^{id}+ f^{ex}.
\end{equation} 
The ideal part of the free energy density is given by a straightforward 
generalisation of the free energy of an ideal gas as
\begin{equation}
\beta f^{id} = \int dR \rho(R) (\ln \rho(R) -1)
\end{equation}
with $\beta = 1/(k_{B}T)$. For the case of the rectangular distribution, the 
ideal free energy density is simply
\begin{equation}
\beta f^{id} = \rho (\ln \rho -1) - \rho \ln(\bar{R}w).
\label{fideal}
\end{equation}
We calculate the excess free energy density, $f^{ex}$, of the polydisperse fluid 
from the general equation of state given by Salacuse and 
Stell\cite{Salacuse-812}. This equation of state is a straightforward 
generalisation of the highly successful EOS for an arbitrary hard sphere 
mixture\cite{Mansoori-852} first considered by Mansoori {\em et al.\/} The 
Mansoori EOS has been checked against simulation data by a number of authors and 
agreement is generally excellent. The free energy per particle (${\cal F} = 
F/N$) is then ${\cal F}_{f} = (f_{f}^{id} + f_{f}^{ex})/\rho$.

Now we consider the free energy of the fractionated solid phase. Since as 
mentioned above, we expect the number of crystals to depend upon the overall 
polydispersity $\sigma$ of the parent distribution, we consider the general case 
of $P$ coexisting crystals. In order to establish some relation between the 
parent distribution $\rho(R)$ and the sub-distribution, $\rho_{i}(R)$, found in 
the $i^{th}$ crystal we make two assumptions. First we choose the distribution 
in each crystal to be rectangular and second, to keep the theory computationally 
manageable, we force the width (or polydispersity) to be equal in each of the 
$P$ coexisting phases. With these two restrictions the housekeeping of the 
fractionation process is straightforward. The $i^{th}$ crystal, for instance, 
contains all $N_{i}$ particles whose diameters lie between $\bar{R}\left[ 1 - 
w_{s}\left(1+\frac{P}{2}-i\right)\right]$ and $\bar{R}\left[ 1 - 
w_{s}\left(\frac{P}{2}-i\right)\right]$ with $w_{s}$, the width of the crystal 
distribution, given by $w_{s} = w/P$. The mean diameter of particles in the 
$i^{th}$ crystal is then simply given as
\begin{equation}
\bar{R}_{i} = \bar{R} \left[ 1 - \frac{w_{s}}{2} \left(1+P-2i\right) \right]
\label{meandiameter}
\end{equation}
and the polydispersity of each crystal is $\sigma_{s} = \sigma / P$. The 
diameter distribution in the $i^{th}$ crystal is then
\begin{equation}
\rho_{i}(R) = \rho_{i} \left\{ \begin{array}{ll}
                        0 & R < \bar{R}_{i}(1-w_{s}/2) \\
                        (\bar{R}_{i}w_{s})^{-1} & \bar{R}_{i}(1-w_{s}/2) \leq R 
\leq    \bar{R}_{i}(1+w_{s}/2) \\
                        0 & R > \bar{R}_{i}(1+w_{s}/2)
				\end{array}
		  \right.
\label{crystal}
\end{equation}
at an overall density of $\rho_{i} = N_{i}/V_{i}$ with $V_{i}$ the volume of the 
$i^{th}$ phase. Although the number of particles in each crystal phase is the 
same for all crystals and equal to $N/P$ (since the polydispersity of each 
crystal is assumed equal) the volume $V_{i}$ of each phase is so far 
undetermined. To fix the volume, or equivalently the number density $\rho_{i}$, 
of each crystal we use the fact that the crystals are in equilibrium with each 
other. This constraint is straightforward to apply since, in our model of the 
partition, no particle may exist in anymore than one crystal. Consequently we do 
not need to consider exchange of particles. Equilibrium simply requires that the 
osmotic pressure of each crystal, $\Pi_{i}$, be fixed and equal to an external 
pressure of, say, $\Pi$
\begin{equation}
\Pi_{i}(\rho_{i},\sigma_{s},\bar{R}_{i}) = \Pi.
\label{constantP}
\end{equation}
With the density of each crystal phase determined, the total number density of 
the set of $P$ coexisting crystals is then
\begin{equation}
\bar{\rho}_{s} = \frac{P}{\sum_{i=1}^{P} (1/\rho_{i})}
\label{effectiverho}
\end{equation}
and the corresponding composite packing fraction is $\bar{\phi} = (\pi/6) 
\bar{\rho}_{s} \bar{R}^{3}(1+w^{2}/4)$. The mean free energy per particle is
\begin{equation}
\bar{{\cal F}}_{s}^{P} = \frac{1}{P} \sum_{i} \frac{f_{i}}{\rho_{i}}
\label{freesolid}
\end{equation}
where $f_{i}$ is the free energy density of the $i^{th}$ crystal. 

To complete our model for the fractionated phases we need an expression for the 
excess free energy of a polydisperse crystal. For that purpose we use a scaled 
particle theory. Since the details have been described 
elsewhere\cite{Bartlett-983} we just outline the approach here. The model 
exploits the picture of particles in a solid as being confined in cells or cages 
formed by the neighbours from which they can not escape. While, in a 
monodisperse crystal the cells are all identical, in a polydisperse crystal the 
size and shape of the cell varies. The idea is to recognise that one can allow 
for the variability in the cell shape, to a first approximation, by considering 
just a finite number of different-sized neighbours. The number of 
different-sized particles is determined by the accuracy with which the 
properties of the finite mixture approximate those of the polydisperse system. 
If the excess free energy of the polydisperse system is a function of the first 
$p$ moments of the diameter distribution $[p/2]$ different-sized spheres are 
needed to match the diameter moments of the continuous distribution ($[x]$ is 
the smallest integer not less than $x$). For a polydisperse system of hard 
spheres, highly successful mean-field theories such as the Percus-Yevick 
approximation suggest that only four diameter moments are 
significant\cite{Salacuse-812}. In this case, the properties of a polydisperse 
system should be well approximated by a binary mixture of spheres, chosen so 
that the first four moments of the polydisperse and binary distributions are 
equal. The two diameter distributions are `equivalent' and all excess properties 
are, to a first approximation, equal. For the rectangular distribution the 
equivalent binary distribution has an equal number of large and small spheres 
with diameters $\bar{R}_{i}(1+\sigma_{s})$ and $\bar{R}_{i}(1-\sigma_{s})$ 
respectively. Specifically then, we equate the excess free energy of the 
polydisperse crystal with that of an equivalent binary substitutional crystal. 
For the evaluation of the latter we take advantage of the analytical expressions 
quoted by Kranendonk and Frenkel\cite{Kranendonk-813} as fits to Monte Carlo 
simulation results. Combining the excess free energy with the ideal free energy 
density,
\begin{equation}
\beta f_{i}^{id} = \rho_{i} (\ln \rho_{i} -1) - \rho_{i} \ln(\bar{R}_{i}w_{s})
\end{equation}
yields the free energy density $f_{i}$ of the $i^{th}$ crystal. Finally, we note 
that our model for the polydisperse crystal implicitly assumes that all 
different-sized spheres are placed randomly on the sites of a common fcc 
lattice. At high polydispersities, the smallest particles in the distribution 
could also be accommodated within the interstices of a crystal of larger 
particles. The current calculations ignore this possibility and so are valid 
only for polydispersities of $\sigma \lesssim 0.3$.

Putting everything together, we now calculate the difference free energy per 
particle, $\Delta {\cal F}^{P} = {\cal F}_{s}^{P} - {\cal F}_{f}$, between the 
polydisperse fluid and the fractionated set of $P$ crystals as follows. The 
overall particle density distribution of either the set of crystals or the 
polydisperse fluid is characterised by identical values for the overall volume 
fraction $\bar{\phi}$, the polydispersity $\sigma$ and the mean diameter 
$\bar{R}$. For fixed $\phi = \bar{\phi}$, $\sigma$ and $\bar{R}$, the free 
energy of the fluid ${\cal F}_{f}$ is calculated from Eqs.\ \ref{fideal} and the 
Mansoori EOS. Combining Eqs.\ \ref{constantP} and \ref{effectiverho} we see that 
the overall volume fraction $\bar{\phi}$ of the solid phases (or equivalently 
the mean density $\bar{\rho}_{s}$) may be considered as a function of the 
osmotic pressure $\Pi$. Inverting this relation yields the osmotic pressure 
$\Pi$ and so, from Eq.\ \ref{constantP}, the densities $\rho_{i}$ of each of the 
coexisting crystals which taken together have an overall volume fraction of 
$\bar{\phi}$. Knowing the density, polydispersity $\sigma_{s} = \sigma /P$ and 
mean diameter $R_{i}$ (from Eq.\ \ref{meandiameter}) of each crystal it is 
straightforward to calculate the corresponding free energy density $f_{i}$ and 
the mean free energy ${\cal F}_{s}^{P}$ of the multiple crystals from Eq.\ 
\ref{freesolid}.

\section{results}
\label{results}

\begin{figure}
\begin{center}
\epsfysize=7.0in 
\leavevmode 
\epsfbox{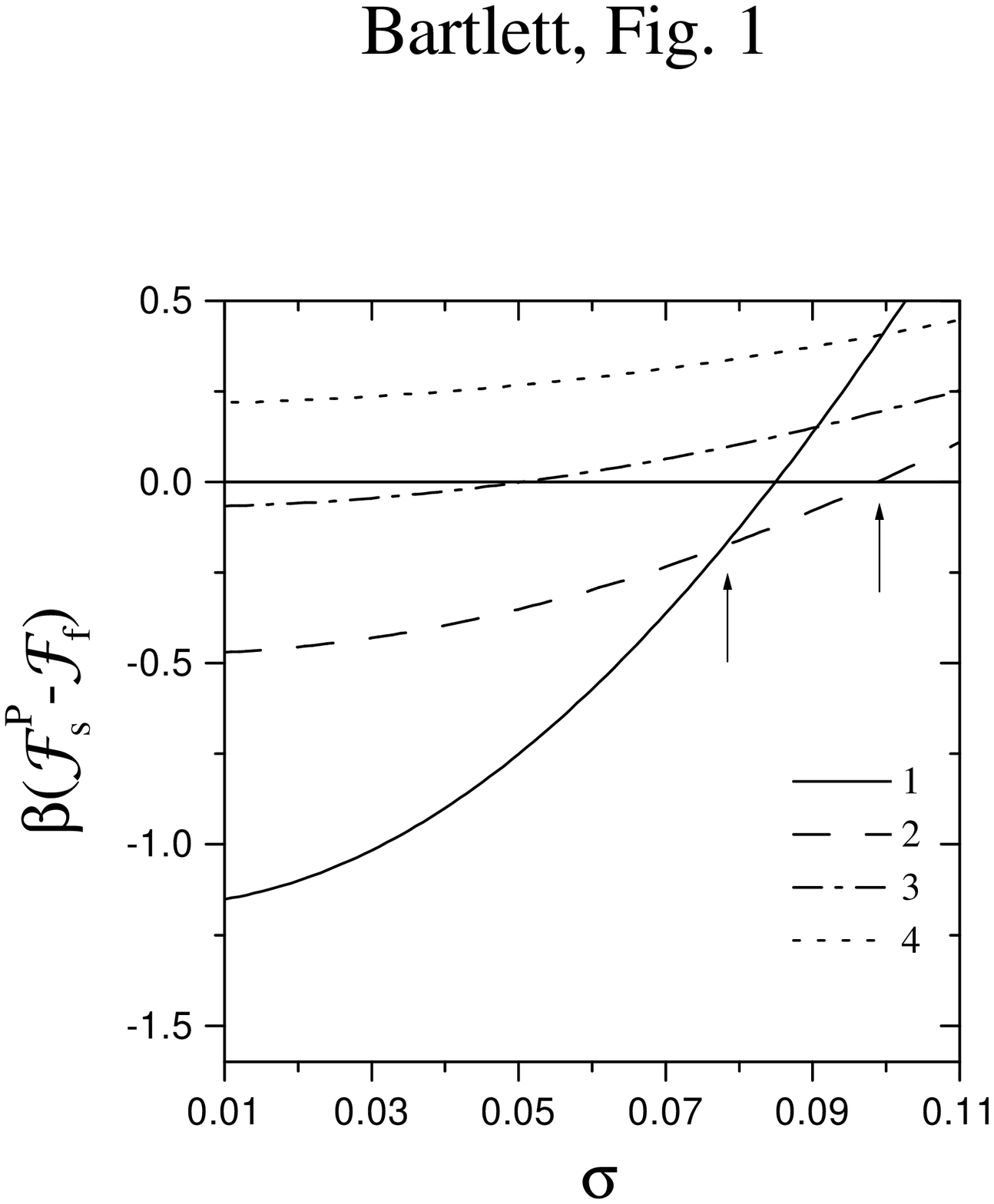}
\end{center}
\caption{The free energy difference per particle, at a constant packing fraction 
of $\phi = 0.6$, between polydisperse crystalline and fluid phases as a function 
of polydispersity. The number of coexisting crystals is detailed in the legend. 
The arrows mark the polydispersity limits between which the fractionated binary 
crystals are stable.}
\label{fig1}
\end{figure}

We now present in detail results for the stability of a system of polydisperse 
hard spheres obtained from the theory described above. In Fig.\ \ref{fig1}, we 
show the free energy difference, $\beta \Delta {\cal F}^{P}$, between the 
fractionated crystals and the fluid phase as a function of polydispersity, for 
the representative density of $\phi = 0.60$. The first question to be addressed 
is the relative stability of the various fractionated crystals and the fluid 
phase.

Referring to Fig.\ \ref{fig1}, we see that for low levels of polydispersity, at 
this density, the crystal has a significantly lower free energy than the fluid 
phase. However, with increasing polydispersity, the stability of the 
polydisperse crystal reduces rapidly until for $\sigma \gtrsim 0.084$ the 
unfractionated polydisperse crystal is unstable relative to the fluid phase. The 
origin of this behaviour lies in the different ways polydispersity affects the 
maximum packing fraction, or equivalently the free volume, of ordered and 
disordered structures. In a fluid, smaller particles are free to pack in the 
cavities between larger particles and so the free volume and thus the entropy 
increases with polydispersity. Conversely, the periodicity of a crystal causes 
the free volume and entropy of a polydisperse crystal to decrease with 
increasing polydispersity. Consequently the free energy of a fixed-density 
polydisperse crystal will diverge at a critical level of polydispersity at which 
the fluid free energy will remain finite. Second, we note from Fig.\ \ref{fig1} 
that at low polydispersities, fractionating the distribution into two or more 
crystals always raises the free energy of the solid state. This is because of 
the loss of entropy of mixing as the diameter distribution is split up. The 
contribution to the free energy difference, $\beta \Delta {\cal F}^{P}$, due to 
the loss of entropy of mixing is simply:
\begin{equation}
\beta \Delta {\cal F}^{P}_{mix} = \ln P
\label{mixing}
\end{equation}
for fractionation into $P$ crystal phases which is very close to the differences 
seen in Fig.\ \ref{fig1} as $\sigma \rightarrow 0$. However at finite 
polydispersities, Eq.\ \ref{mixing}, does not give a reasonable estimate of the 
free energy differences between the various sets of crystals. As is evident, the 
free energy $\beta \Delta {\cal F}^{P}$ reduces with increasing polydispersity. 
The reduction being smaller as the number of crystals considered increases. This 
is because the polydispersity of each fractionated crystal drops as the total 
number of crystals, between which the overall distribution is split, increases. 
For instance, fractionating a distribution with a polydispersity of 6\% between 
two phases would result in crystals with 3\% polydispersity, three phases would 
lead to 2\% polydispersity and so on. As the polydispersity of each of the 
individual crystals decreases the influence of the divergence of the free energy 
at the close-packing limit is reduced and so the free energy is less affected by 
the overall polydispersity. As a consequence the fractionated two-crystal phase 
takes the place of the polydisperse single-phase crystal, as the state of lowest 
free energy, for $\sigma > 0.078$ and remains the most stable phase until 
$\sigma = 0.099$, at which point the fluid appears.

By a procedure similar to that described above, we have mapped out the range of 
stability of the polydisperse hard sphere system for $\sigma \leq 0.20$. Our 
results are shown in Fig.\ \ref{fig2}. The regions are labelled by the number of 
coexisting crystals which minimise the total free energy of the polydisperse 
system while the boundaries indicate the specific densities and polydispersities 
at which competing phases have equal free energies. As is evident, there are 
appreciable areas of density and polydispersity where fractionation into 
multiple crystals is predicted. 

At low volume fractions, $\phi < 0.516$, the polydisperse fluid is the most 
stable phase at all polydispersities. However with an increase in packing 
fraction a crystalline phase takes the place of the polydisperse fluid as the 
most stable phase. The nature of the crystalline phase depends on the magnitude 
of the polydispersity $\sigma$. So while for $\sigma < 0.085$ we have 
crystallisation into the usual single polydisperse crystal, with increasing 
polydispersity the fluid fractionates upon solidification into a rapidly 
increasing number of crystals. The diagram gets more and more complicated as the 
polydispersity grows since, by necessity, the number of crystalline phases 
required to accommodate the increasing width of the diameter distribution 
increase. So while for polydispersities in the range $0.085 < \sigma < 0.117$ 
two fractionated crystals are first formed, three crystals are required for 
$0.117 < \sigma < 0.145$, four for $0.145 < \sigma < 0.171$, five for $0.171 < 
\sigma < 0.195$ while at $\sigma = 0.20$ complete crystallisation requires six 
different crystal phases. Furthermore, we note that as the polydispersity 
increases the packing fraction at which the fluid becomes unstable increases 
from $\phi = 0.516$ at $\sigma = 0$ to $\phi= 0.665$ at $\sigma = 0.20$. The 
degree of fractionation predicted varies with the overall level of 
polydispersity. Referring to Fig.\ \ref{fig3}, we see that the polydispersity of 
the crystal(s) at the fluid-solid boundary reaches a maximum of 0.085 before on 
average falling with increasing fluid polydispersity, although in a far from 
continuous fashion. Where fractionation occurs, Fig.\ \ref{fig3} reveals that 
the polydispersity of each crystal formed is considerably lower than the 
critical polydispersity, estimated here as $\sigma_{c} \approx 0.085$. 

In addition to the fluid-multiple crystal transitions discussed above, Fig.\ 
\ref{fig2} reveals that there is a further sequence of demixing transitions in 
the solid phase. Compressing a polydisperse crystal causes a solid-state phase 
separation in which the diameter distribution is fractionated. For instance a 
polydisperse crystal with $\sigma = 0.08$ is stable up to a density of $\phi = 
0.597$. At this point the system separates into two crystals, each of 
polydispersity $\sigma = 0.04$, which are stable until a total density of $\phi 
= 0.667$ is reached. Further compression results in three crystals, followed at 
still high densities by a cascade of demixing transitions. The explanation for 
this behaviour lies in the effect of polydispersity on the limiting packing 
fraction of a crystal. As mentioned previously, the density at which the free 
energy of a polydisperse crystal starts to diverge reduces with increasing 
polydispersity. As a polydisperse crystal is compressed there comes a point at 
which the reduction in the excess free energy which occurs when the 
polydispersity is reduced is sufficient to exceed the increase in the ideal free 
energy of mixing after fractionation. At this point the crystal phase separates.

Finally, we emphasise that we have calculated only the stability boundaries, not 
the full equilibrium phase diagram of a system of polydisperse spheres. The 
reason for this is that a full calculation of the phase equilibria, particularly 
in view of the large number of competing phases found in this work, would be a 
large and complicated problem. However we expect our stability diagram (Fig.\ 
\ref{fig2}) to be a useful guide to the form of the phase diagram. Confirmation 
of this point of view is provided by the comparison, seen in Fig.\ \ref{fig4}, 
between the current model and the fluid-single crystal coexistence region 
established in earlier work\cite{Bartlett-1266}. The dashed curves indicate the 
positions of the polydisperse cloud-point boundaries and were calculated by a 
novel moment projection method using, as input, polydisperse free energies 
similar to those described above. Referring to Fig.\ \ref{fig4}, we see that the 
stability boundary lies approximately midway between the two coexistence 
densities. Inspection of Fig.\ \ref{fig4} reveals, rather intriguingly, that the 
polydisperse point of equal concentration, marked by the large circle in Fig.\ 
\ref{fig4}, is practically coincident with the lowest density at which the 
binary system of fractionated crystals is stable. This seems to be a chance 
occurrence. However, the proximity of the two points suggests that, at 
equilibrium, the re-entrant melting predicted in Ref. \cite{Bartlett-1266} will 
in fact be pre-empted by a solid state phase separation.

\section{Discussion and Conclusions}
\label{discussion}

We have analysed the stability of a polydisperse fluid of hard spheres with 
respect to a process of simultaneous fractionation and crystallisation. Although 
the model is quite simple we find several interesting features. In particular, 
we predict that for a sufficiently polydisperse system of hard spheres 
fractionation occurs upon solidification. The fluid diameter distribution 
separates into a number of fractions of narrower polydispersity which then 
crystallise. Furthermore, we find that compressing a polydisperse crystal 
induces a sequence of demixing transitions in the crystal.

To keep the theory presented here manageable we have been forced to make a 
number of assumptions. We have, for instance, imposed a somewhat arbitrary model 
for the fractionation process. In particular we have assumed that a rectangular 
diameter distribution fractionates into a number of equal-width daughter 
distributions, each of which is also rectangular in form. In principle, we could 
relax the constraint of equal width. However, since the densities of the 
coexisting crystals are similar it seems likely that this modification would 
have little effect. The more critical restriction is probably the assumption 
made for the form of the daughter distribution. However, in view of the relative 
insensitivity of polydisperse systems to the detailed form chosen for the 
diameter distribution, we expect that the current predictions will be in at 
least qualitative if not quantitative agreement with more rigorous calculations, 
when available. 

We now speculate briefly on the feasibility of observing solid-state 
fractionation in experiments on hard-sphere colloids. The first point to note is 
that fractionation requires colloid diffusion over distances comparable to the 
size of the growing crystallite. The rate of such large scale diffusive motion 
reduces with increasing colloid density, as a result of caging effects, and 
essentially vanishes at the glass transition density, 
$\phi_{g}$\cite{Pusey-853}. For uniformly-sized hard spheres 
simulations\cite{Woodcock-87} find a long-lived metastable glassy state around 
$\phi_{g} = 0.58$ which persists up to the random close packing limit at 
$\phi_{rcp} = 0.64$. The glass transition in a system of polydisperse hard 
spheres has not been studied to the best of our knowledge although Schaertl and 
Silescu\cite{Schaertl-828} have estimated $\phi_{rcp}$ as a function of 
polydispersity from simulation. Assuming a simple linear relationship between 
the two densities gives a crude estimate of the glass transition density as 
$\phi_{g}(\sigma) \approx (0.58/0.64)\phi_{rcp}(\sigma)$. Referring to Fig.\ 
\ref{fig4}, we see that if this estimate for the glass transition is used, then 
fractionated crystallisation might be experimentally observable for 
polydispersities in the narrow range $0.08 < \sigma < 0.11$ where diffusive 
motion is still possible although slow. If conversely, the dynamics of the 
fractionation process turn out to be appreciably slower than the experimental 
timeframe then experiments should follow our earlier 
predictions\cite{Bartlett-1266} and exhibit a polydisperse point of equal 
concentration. A clear resolution of this ambiguity must await experiment.

\begin{figure}
\begin{center}
\epsfysize=7.0in 
\leavevmode 
\epsfbox{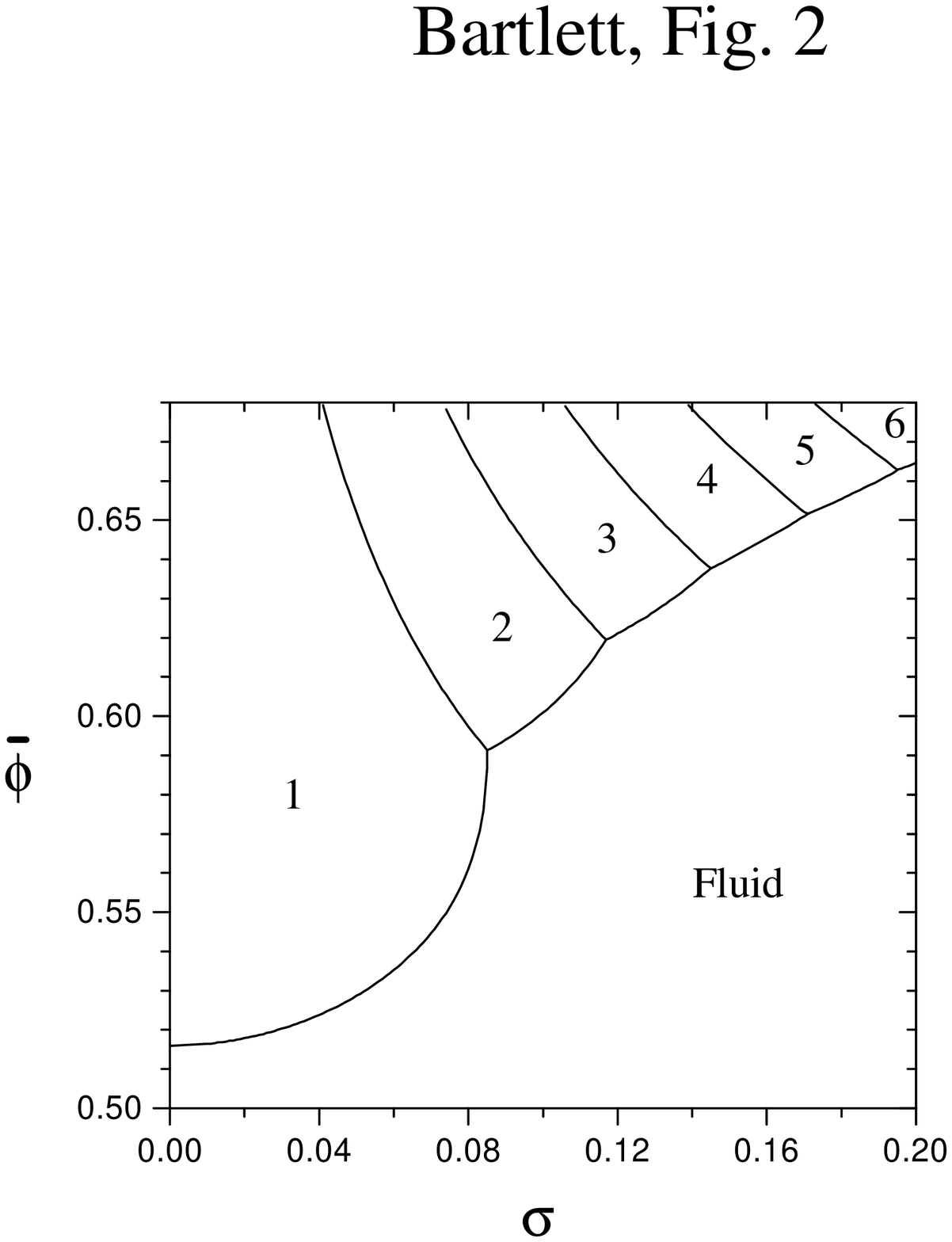}
\end{center}
\caption{The stability diagram of a polydisperse hard sphere mixture. In each 
region of volume fraction $\bar{\phi}$ and polydispersity $\sigma$ the phase 
with the lowest free energy is indicated. Crystalline regions are labelled by 
the number of coexisting crystal phases.}
\label{fig2}
\end{figure}

\begin{figure}
\begin{center}
\epsfysize=7.0in 
\leavevmode 
\epsfbox{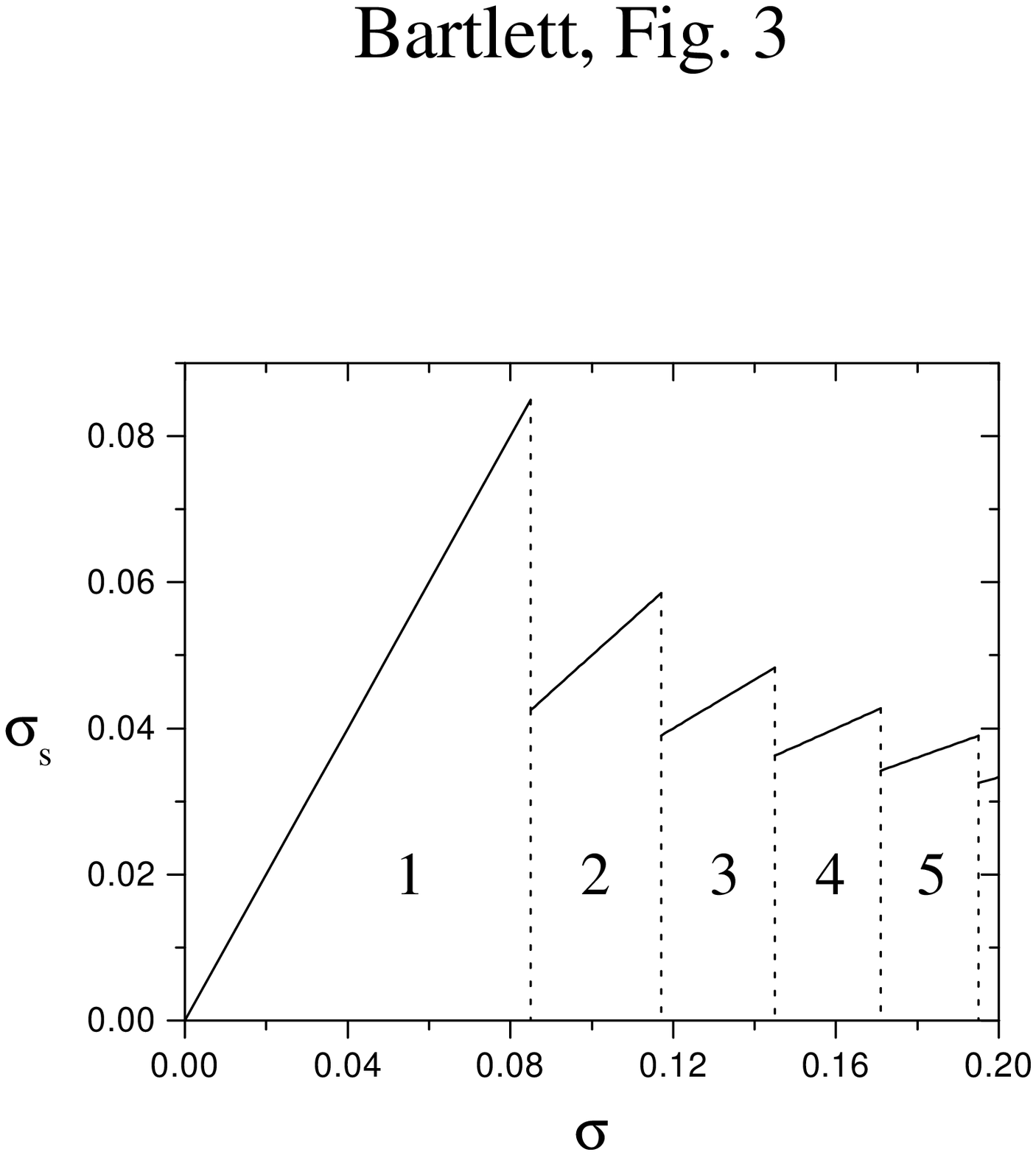}
\end{center}
\caption{The polydispersity, $\sigma_{s}$, of the fractionated crystalline phase 
formed at the fluid-solid boundary as a function of the fluid polydispersity 
$\sigma$.}
\label{fig3}
\end{figure}

\begin{figure}
\begin{center}
\epsfysize=7.0in 
\leavevmode 
\epsfbox{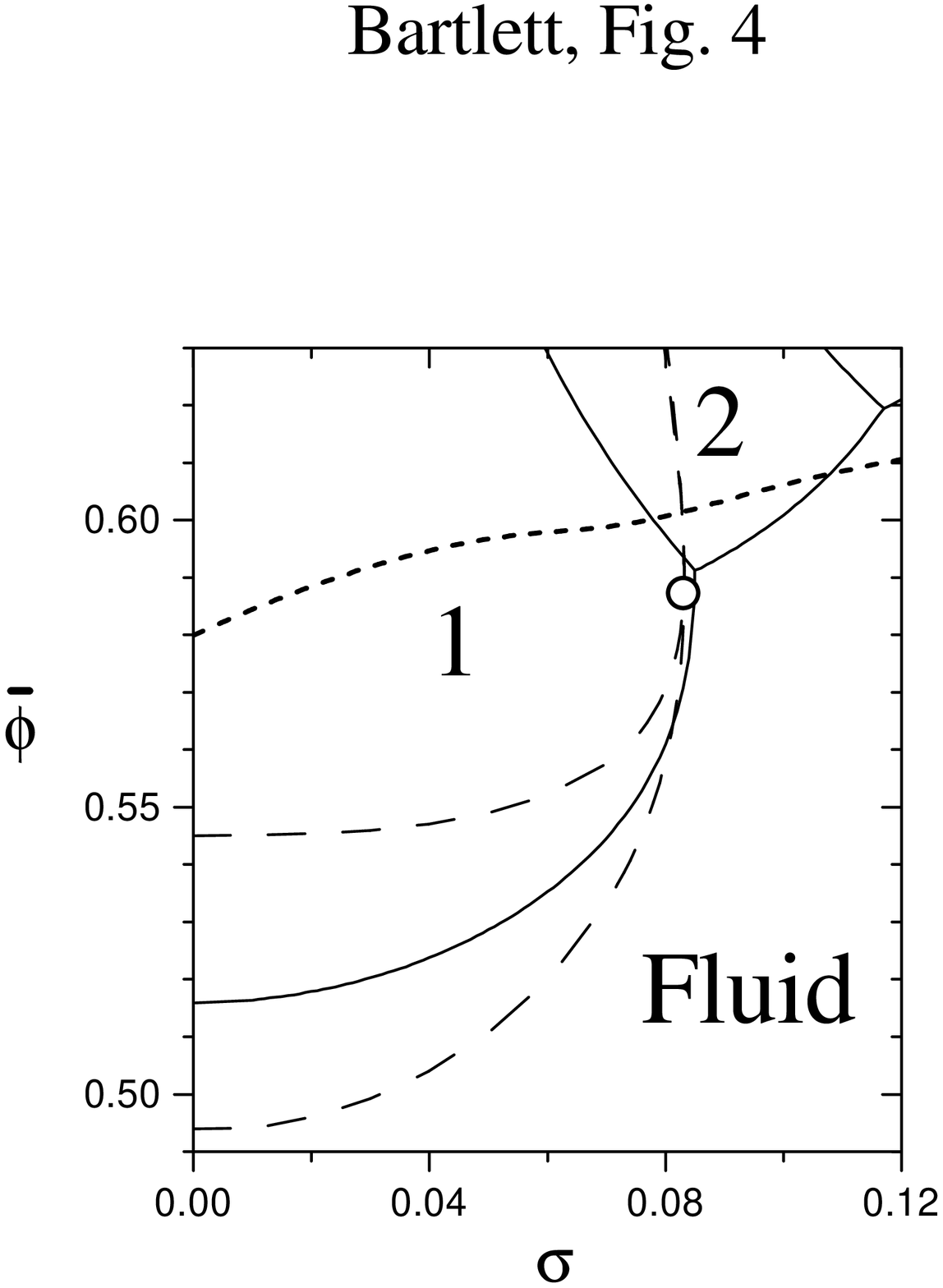}
\end{center}
\caption{A comparison between the current stability diagram (solid lines) and 
the fluid-solid coexistence predicted in Ref. \protect{\cite{Bartlett-1266}} 
(dashed lines). The large circle marks the position of the polydisperse point of 
equal concentration. The thick  dotted line denotes the estimated location of 
the polydisperse glass transition.}
\label{fig4}
\end{figure}

Finally, our results suggest a possible explanation for the apparently anomalous 
simulations reported by Bolhuis and Kofke\cite{Bolhuis-824}. In contrast with 
other theoretical work, this study found that: (i) the particle size 
distributions in the fluid and crystal phases were significantly different  and 
(ii) that the coexistence region although initially narrowing with increasing 
polydispersity finally widened. The phase diagram was traced out by integrating 
along the fluid-crystal coexistence line. Implicit in this approach is the 
assumption that only one crystalline phase exists at the fluid-solid boundary. 
Referring to Fig.\ \ref{fig2}, we envisage that at the point where two 
polydisperse crystals become stable the simulation will follow one of the two 
fluid-solid branches until it halts at the discontinuity marking the start of 
the three-crystal region. With this picture in mind, we may readily account for 
the observation that at the point, where the simulation could no longer follow 
the fluid-solid boundary, the crystal polydispersity was almost exactly half 
that of the fluid value. At the same time, we note that their limiting 
polydispersity ($\sigma = 0.118$) lies close to our estimate for the highest 
polydispersity (0.117) at which two polydisperse crystals remain the equilibrium 
phase.

%
%
\end{document}